# Rogue events in complex linear and nonlinear photonic media

M. Mattheakis[1,2,3,*], I. J. Pitsios[1,4,*], G. P. Tsironis[1,2,5], S. Tzortzakis[1,4,6]

1. Institute of Electronic Structure and Laser, Foundation for Research and Technology Hellas, P.O. Box 1527, 71110, Heraklion, Greece
2. Physics Department, University of Crete, P. O. Box 2208, 71003,Heraklion, Crete, Greece
3. School of Engineering and Applied Sciences, Harvard University, Cambridge, Massachusetts 02138, USA
4. Materials Science and Technology Department, University of Crete, P.O. Box 2208, 71003, Heraklion, Greece
5. National University of Science and Technology MISiS, Leninsky prosp. 4, Moscow, 119049, Russia
6. Science Program, Texas A&M University at Qatar, P.O. Box 23874 Doha, Qatar

*These authors contributed equally to this work

**Ocean rogue waves (RW) –huge solitary waves– have for long triggered the interest of scientists. RWs emerge in a complex environment and it is still dubious the importance of linear versus nonlinear processes. Recent works have demonstrated that RWs appear in various other physical systems such as microwaves, nonlinear crystals, cold atoms, etc. In this work we investigate optical wave propagation in strongly scattering random lattices embedded in the bulk of transparent glasses. In the linear regime we observe the appearance of RWs that depend solely on the scattering properties of the medium. Interestingly, the addition of nonlinearity does not modify the RW statistics, while as the nonlinearities are increased multiple-filamentation and intensity clamping destroy the RW statistics. Numerical simulations agree nicely with the experimental findings and altogether prove that optical rogue waves are generated through the linear strong scattering in such complex environments.**

Ocean rogue or freak waves are huge waves that appear in relatively calm seas in a very unpredictable way. Numerous naval disasters leading to ship disappearance under uncertain conditions have been attributed to these waves. Since sailors are well known story makers these monster, destructive waves that were in naval folklore perhaps for thousands of years penetrated the realm of science only recently and after quantitative observations [1,2]. Since then, they seem to spring up in many other fields including optics [3-7], BEC and matter waves, finance, etc [8-12]. Unique features of rogue waves, contrary to other solitary waves, are both their extreme magnitude but also their sudden appearance and disappearance. In this regard they are more similar to transient breather events than solitons. Since the onset of both necessitates the presence of some form of nonlinearity in the equation of motion describing wave propagation, it has been tacitly assumed that extreme waves are due to nonlinearity. Intuitively, one may link the onset of a rogue wave to a resonant interaction of two or three solitary waves that may appear in the medium. However, large amplitude events may also appear in a purely linear regime [1,2,4,6]; a typical example is the generation of caustic surfaces in wave propagation [13,14].

Propagation of electrons or light in a weakly scattering medium is a well-studied classical problem related to Anderson localization and caustic formation. Recent experiments in the optical regime [15] have shown clearly both the theoretically predicted light localization features as well as the localizing role of (focusing) nonlinearity in the propagation [15-20]. In these experiments a small (of the order of $10^{-3}$) random variation of the index of refraction in the propagation leads to eventual localization while at higher powers, where nonlinearity is significant, localization is even stronger. Thus, destructive wave interference due to disorder leads to Anderson localization that may be enhanced by self-focusing nonlinearity. In the purely linear regime propagation in two dimensions in a weakly random medium has shown that branching effects appear through the generation of caustic surfaces [13, 14], while linear rogue waves have been observed with microwaves [4].

In this work we focus on an entirely different regime of wave propagation, in strongly scattering optical media that consist of Luneburg-type lenses randomly embedded in the bulk of glasses. Spherical or cylindrical Luneburg lenses (LLs) have very strong focusing properties directing all parallel rays impinging on them to a single spot on the opposite side surface. The index variation is very large, viz. of the order of 40% and thus a medium with a random distribution of Luneburg-type lenses departs strongly from the Anderson regime investigated in [15-20]. In the experimental configuration used in this work we used "Luneburg Holes (LH)" or anti-Luneburg lenses instead of LLs; the LHs have a purely defocusing property. In the methods section we demonstrate that our observations discussed in the following are generic and independent of the type of scatterers.

**Experimental and numerical observation of rogue waves.** Focusing tightly a femtosecond IR beam into the bulk of fused silica substrates induces nonlinear absorption allowing the selective modification of the material [21]. Under appropriate irradiation conditions one may create LH-type structures and by placing those in a controlled way in space to create three dimensional LH lattices like the ones shown in Fig. 1(a).

The investigation for the presence of a rogue wave is performed by probing a laser beam through the volume of the lattice and imaging the output. This approach is advantageous because it allows the study of both linear and nonlinear phenomena, depending only on the probe beam intensity.

For the linear observations a low power continuous wave 633 nm laser beam was used as probe. A large number of different lattices were studied until "rogue" events were observed as seen in Fig. 1(b). The corresponding "rogue" event intensities profile is shown in Fig. 1(c) and the distribution of the intensities, in semilog scaling, in Fig. 1(d) and permit to conclude that this signal cannot be anything else than an optical rogue wave, contiguous to the definition of the phenomenon [1,2,4].

*Simulations*

In Fig. 2 we present the light propagation in a random LH lattice (Fig. 2a) under steady state conditions. We observe that the presence of scatterers with strong defocusing properties forces light to form propagation channels (Fig. 2b) that can lead in the generation of very large amplitude rogue wave events (Fig. 2c). Such events have amplitudes larger than twice the significant wave height (SWH) in the medium and are directly attributed to wave coalescence induced by the strong scattering of light by the LHs. Although the medium is purely linear, the induction of caustic surfaces leads to resonant events that have clear rogue wave signatures. In Fig. 2(d) is shown the intensity profile where a rogue wave occurs. Obviously, the highest peak is larger than twice the SWH resulting in a rogue wave event.

In Fig. 2(e) is plotted the distribution of intensities (in semilog axis). By the central limit theorem and the simple random wave prediction for the probability distribution of wave intensities I, the intensities have to follow the Rayleigh law, meaning a distribution type $P(I) = e^{-I}$ where $I = |E|^2$ (E is the electric field) is normalized to one [1,2,4,14,18]. However, when extreme events appear, the intensities distribution deviates from simple exponential and long tails appear [1,2,4,8], clearly seen in both our experimental Fig. 1(d) and numerical results Fig. 2(e).

**Experimental and numerical parametric studies:** In order to study the dependence of the phenomenon on the scattering strength of the lattice we vary the LH lattice randomness as well as the refractive index profile amplitude. This is done by fabricating various system configurations with different distribution of LHs as well as different maximal $\Delta n$ differences in the index of refraction between the host medium (glass) and the center of the LH. Interestingly changing the disorder level did not alter the general RW statistics picture. On the other hand, the refractive index variation of the sample influenced the phenomenon strongly. Specifically we found that there is a threshold in $\Delta n$ below which no rogue waves were observed. The variation from small values ($\Delta n < 1\%$) Fig. 3(a), to intermediate values ($\Delta n \approx$ few %) Fig. 3(b) and high values ($\Delta n \approx 30\%$) Fig. 3(c) shows the clear dependence of the rogue wave generation on the scattering properties of the medium.

In our numerical analysis we investigate random lattices of the type shown in Fig. 2(a) while changing the maximal index variation. In Fig. 4 we present the distribution of intensities for three different index variations, viz. $\Delta n = 10\%$ Fig. 4(a), $\Delta n = 20\%$ Fig. 4(b) and $\Delta n = 30\%$ Fig. 4(c). We found that the long tails at high intensities disappear as the index variation decreases, with rogue events appearing for index variations roughly above 20%. The qualitative as well as quantitative agreement of experimental and theoretical results in the linear regime demonstrates that in the present context the onset of RW extreme events is due to strong scattering in the complex LH lattice.

**The role of nonlinearity:** An obvious question arises as of the role of nonlinearity in the same processes. For answering this question experimentally we increased the intensity of the probing radiation (using high power femtosecond pulses) exciting thus nonlinear modes through Kerr nonlinearity. In Fig. 5 one can observe the total beam, Kerr-induced, self-focusing in the bulk of a glass without any lattice inscribed in it as the input beam power is increased from (a) to (d). On the contrary when the same intense beam goes through a glass with a lattice inscribed in it things are considerably different. At the limit of small nonlinearity, around the critical power, although an amplification of the waves already existing in the linear regime is observed, the linear RW statistics are not modified. This is shown in Fig. 6(a) where a linear RW is further amplified maintaining though its intensity aspect ratio compared to the neighboring lower level waves. As the input power is increased gradually, the lower height waves are amplified as well resulting to a small amplitude multi-filamentation image, Fig 6(b). Further increase in the input beam power, and thus higher nonlinearity, results to the saturation of the intensity of all modes, starting from the higher to the lower ones, since higher order –defocusing– nonlinearities lead to intensity clamping [22]. This is shown in Fig. 6(c) where a higher input laser power pushes many small waves up to the clamping intensity. From the above it is clear that the generation of RWs in the strongly scattering system is a result of linear interference mechanisms while nonlinearity will either accentuate the phenomenon, when it is relatively small, or completely destroy the RW statistics when it is high. It is interesting to refer here at a recent report on laser filamentation merging and RW events [23]. Actually, these observations can be nicely explained in the frame of our present findings, since the merging of the filaments (although a nonlinear effect) happens in a rogue way not because of the nonlinearity but because of linear thermal effects and turbulence induced in the medium by the accumulated heat from the high repetition rate and power of the employed laser system.

Further, our experimental findings on the nonlinearity role are nicely reproduced by numerical simulations (Fig. 7). We introduce a focusing nonlinearity (Kerr effect) in the dielectric constant reading $\varepsilon = n^2 = \varepsilon_L + \chi |\mathrm{E}|^2$, where $E$ is the electric field, $\varepsilon_L$ the linear part of the dielectric constant and $\chi$ the nonlinear parameter varying from 0

up to $10^{-5}$ (depending on the strength of the nonlinearity; in normalized values). As in the experiments we can see that the linear observed RW statistics (Fig.7a,e) are not affected in the presence of a relatively small nonlinearity (Fig. 7b,f). In this case most waves are simply amplified without destroying the RW statistics but slightly increasing the queue of the intensity distribution (Fig. 7f) as expected from the higher amplitudes. This situation dramatically changes at higher nonlinearities (Fig. 7c,g and Fig. 7d,h) where more and more waves are amplified, completely destroying the rogue wave statistics, in full agreement with our experimental observations.

*Conclusion*

Rogue waves are extreme waves that appear in diverse systems; we focused on complex media where randomly placed elements introduce strong light scattering and interference patterns. In the purely linear regime the coalescence of these light channels and the resulting complexity leads to the appearance of extreme, transient waves. There is a clear departure from the Rayleigh law in large intensities where RWs are produced. Most importantly we have shown both experimentally and numerically that the medium nonlinearity does not destroy the RW statistics but rather enhances events that are nucleated in the linear regime. Nevertheless, at higher nonlinearities, the RW statistics are destroyed since many small waves are amplified to large clamped amplitudes. Thus, we conclude that optical extreme events in scattering media are generated by the complexity of the medium that drives interference and wave coalescence. These findings although specific to optical scattering systems, could have direct implications in other physical systems where "scattering" or turbulent effects may be present.

**Methods:**

*Experimental*

The inscription of the scatterers is performed via laser induced refractive index modification. A pulsed IR laser beam (pulse duration 30 fs, central wavelength 800 nm) is focused tightly with an objective lens (x20, NA 0.45). The intensity at the focal volume of the objective lens is high enough to excite nonlinear phenomena such as, nonlinear absorption and avalanche ionization, which in turn can alter permanently the refractive index of optically transparent solid materials, like silica glass [21, 24]. In order to inscribe a scatterers lattice, a glass substrate is mounted on a computer controlled system of translation stages, which allows to move freely in all three dimensions. This approach enables to fabricate different scatterer configurations by changing their coordinates and also facilitates the control of the refractive index variation by simply tuning the radiation intensity and the exposure time per site.

For creating high contrast LHs we used enough energy per pulse to create small voids at the focus [25] resulting thus to peak index changes up to ~0.5. Using third harmonic generation microscopy we found that our LH have a prolate spheroid shape with dimensions of 2-3 μm for the transversal semi-principal axes while the longitudinal are approximately 8-10 μm. The lattices inscribed consist of five superposed layers of 400 LH for each layer. The size of each layer is 250x250 μm and the separation between each layer is 20 μm (Fig. 1a). The size and the density of each lattice were carefully chosen to avoid any overlapping between the lenses.

In order to image a rogue wave, a laser beam is probed through the lattice with its propagation axis (z axis) perpendicular to the layers' plane. An imaging system collects the information at the output (exit layer). The imaging system is mounted on a translation stage in order to be able to image different planes along the propagation. For the studies in the linear regime a 633 nm CW beam is used, while for the non-linear studies the beam comes from the same intense fs laser used for the inscription process.

*Simulations*

The numerical simulations have been performed by solving numerically the full time dependent Maxwell equation using the Finite Difference in Time Domain method (FDTD) [26]. We considered an optical medium with permeability $\mu=1$ and permittivity $\varepsilon(r)=n(r)^2$. Additionally we also employ and reconfirm these results with a ray tracing method using Hamiltonian ray optics [28]. The equivalent system we consider consists of a two dimensional rectangular dielectric medium with uniform index of refraction $n_0=\sqrt{2}$ with embedded cylindrical Luneburg holes (LH) with refractive index $n(r)=\sqrt{1+(r/R)^2}$, where R is the radius of the lens and r denotes the distance from the center of the lens in its interior. The LH's are placed randomly in the medium using a self-avoiding random walk procedure for their centers. This medium is illuminated with TM polarized monochromatic EM plane waves of wavelength λ along the x axis and we follow light propagation on the z-axis; for the theoretical work we use as unit of length the wavelength λ. The medium has dimensions (175.0 x 528.5) (in $\lambda^2$ units), while the radius of each LH is R=3.5λ; we use 400 lenses and place them randomly in the dielectric with a filling factor f=0.17 and absorbing boundary conditions that simulate best the experimental conditions.

In order to control the maximal index variation, which is used in the parametric studies section, we generalized the LH refractive index by introducing a "strength" parameter α according to the expression $n(r)=\sqrt{n_{LH}^2+(1-\alpha)(1-(r/R)^2)}$ where $n_{LH}$ is the original LH index. It can be seen that for α=1 the refractive index reduces to the original LH index, however for α=0 we obtain a flat index with rate equal to $\sqrt{2}$ and zero variation $\Delta n=0\%$.

To confirm the broader validity of our results we investigated numerically the propagation of EM plane waves through other random systems with different scatterer types. We considered two additional cases, one system consisting of LH with constant refractive index (flat-LH) and another system consisting of lenses which have index profiles closer to the experimental ones (exp-LH), where the small index increase at the borders is the result of the overpressure of the material pushed away from the central void. In Fig. 8 we present COMSOL simulations for the EM plane wave propagation through single flat-LH and exp-LH (Figs. 8a, 8c) as well as their corresponding refractive index profiles (Figs. 8b, 8d). In Fig. 9 we show that RWs exist also in the networks consisting of both flat-LH and exp-LH, generalizing thus our findings and concluding that RWs appear in strongly scattering random networks independently of the type of the scatterers.

Although all the simulations presented here are in 2D geometries we have confirmed the validity of these findings performing also simulations in 3D. Because of the unrealistically long time needed to perform high resolution 3D simulations, these were performed only in low resolution and thus, since the findings are very similar, we have decided to present here only the high resolution 2D results.

**Acknowledgements**

We gratefully acknowledge the assistance of M. Thévenet and D. Gray at the early experimental stages of this study. This work was supported by the THALES projects "ANEMOS" and "MACOMSYS", and the Aristeia project "FTERA" (grant no 2570), co-financed by the European Union and Greek National Funds. We also acknowledge partial support through the European Union program FP7-REGPOT-2012-2013-1 under grant agreement 316165 and partial support of the Ministry of Education and Science of the Russian Federation in the framework of increased competitiveness program of MUST "MISiS" ( K2-2015-007).

**Author Contributions**

All authors have contributed to the development and/or implementation of the concept. G.P.T. proposed the original theoretical system with the LLs, while S.T. suggested the experimental realisation with the LHs. S.T. designed and supervised the experimental work and I.J.P. performed the experiments and analysis. M.M. preformed the simulations and analysis, while G.P.T. supervised the theoretical work. All authors contributed to the discussion of the results and to the writing of the manuscript.

**Additional information**

Competing financial interests: The authors declare no competing financial interests.

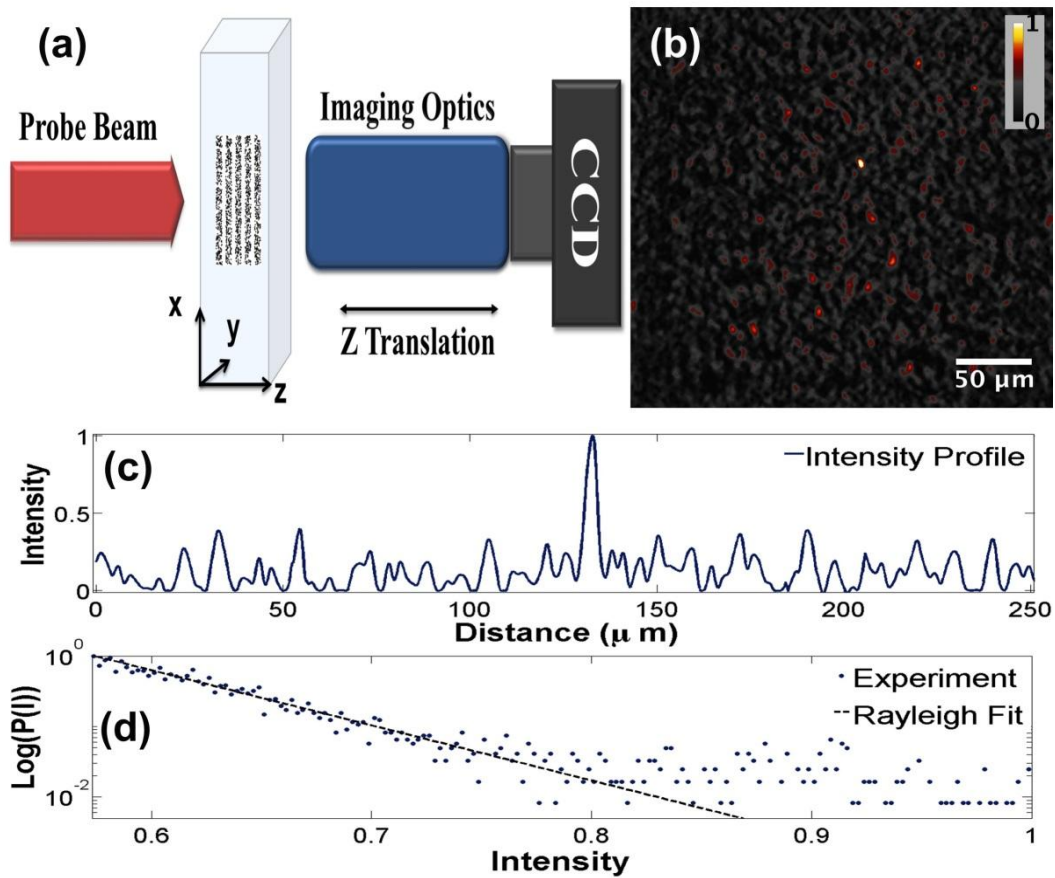

**Figure 1:** Experiments: (a) Schematic representation of the experimental setup. A monochromatic coherent plane wave laser beam propagates from the left to right (red arrow) in the glass sample where a five layer random LHs lattice is inscribed. An imaging system allows recording the beam profile at various propagation planes. (b) Experimental observation of an optical rogue wave as it is formed within the LHs lattice (appearing at the 4[th] layer; almost at the center of the image). The RW is clearly distinct as its intensity is significantly greater from every other wave in the surrounding area in the lattice as seen also at the corresponding intensity profile (c). (d) Intensities distribution (in semilog scaling); rogue waves presence introduces a substantial deviation from the exponential distribution (Rayleigh law) appearing as a tail at high intensities.
10

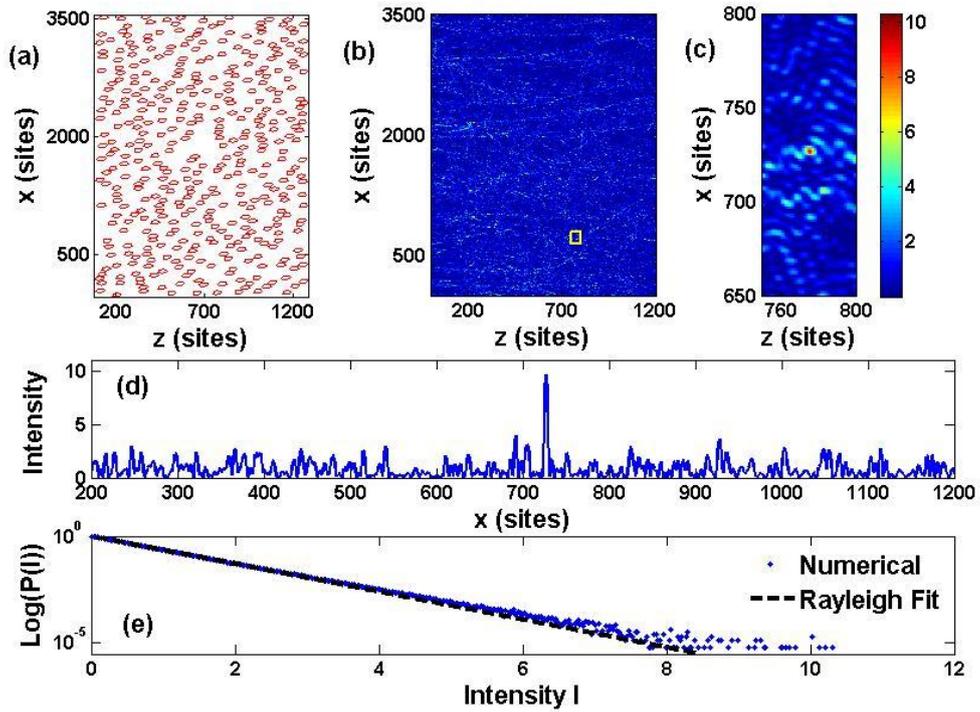

**Figure 2:** Simulations: (a) A 2D random LHs network used in the simulations; each red circle represents a LH. (b) A monochromatic plane wave beam (along x) propagates from left to right (along z) through the lattice. (c) A detail of the propagation (box in (b)) showing an optical RW. (d) Intensity profile in the RW region as a function of x. (e) Intensities distribution (in semilog scaling) for the whole lattice.

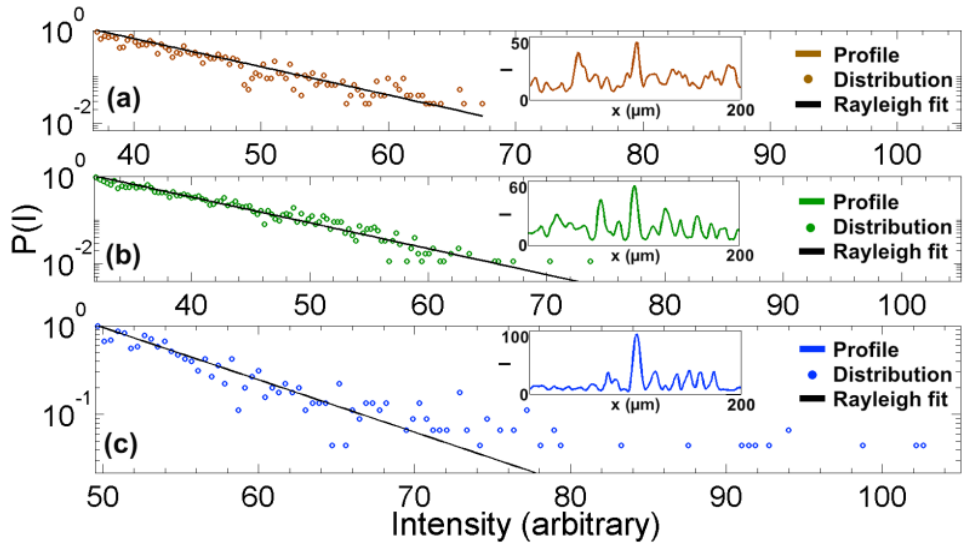

**Figure 3:** Experimental dependence of RW generation on the refractive index amplitude variation $\Delta n$. Intensity distributions and characteristic corresponding profiles (inserts) are plotted for three lattices with different refractive index variations ($\Delta n$): (a) $\Delta n < 1\%$, (b) $\Delta n \approx$ few % and (c) $\Delta n \approx 30\%$. The results show that only for large $\Delta n$ (c) there are strong deviations from the Rayleigh distribution that is accompanied by RW generation.

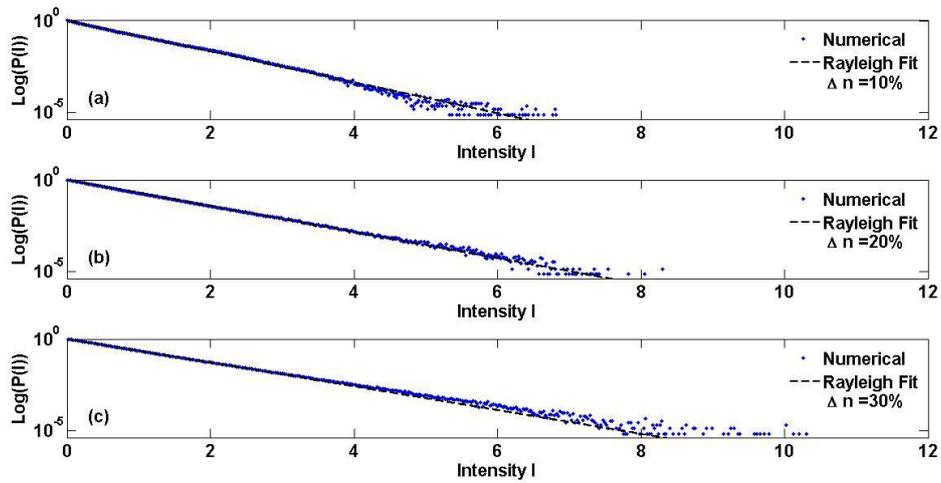

**Figure 4:** Simulation results on the dependence of RW generation on the refractive index amplitude variation $\Delta n$. Intensity distributions are plotted for three lattices with (a) $\Delta n = 10\%$, (b) $\Delta n = 20\%$ and (c) $\Delta n = 30\%$. As in the experiments only for large $\Delta n$ (c) there are strong deviations from the Rayleigh distribution that is accompanied by RW generation.

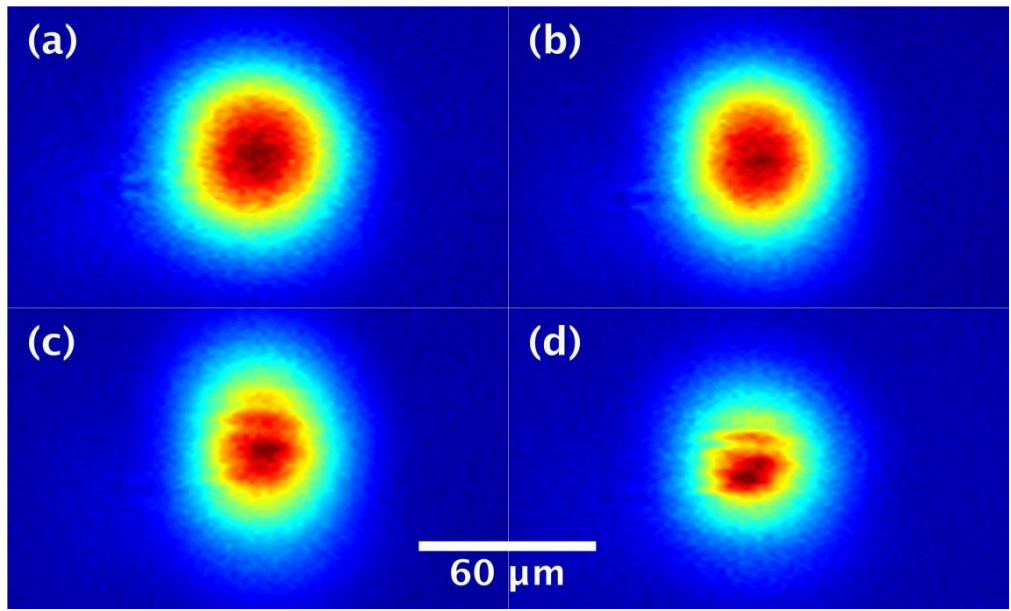

**Figure 5:** Experimental results on the nonlinear propagation of an intense femtosecond probe beam in the bulk of a glass without any lattice. The total beam Kerr self-focusing can be clearly seen as the input laser power is increased from (a) to (d).

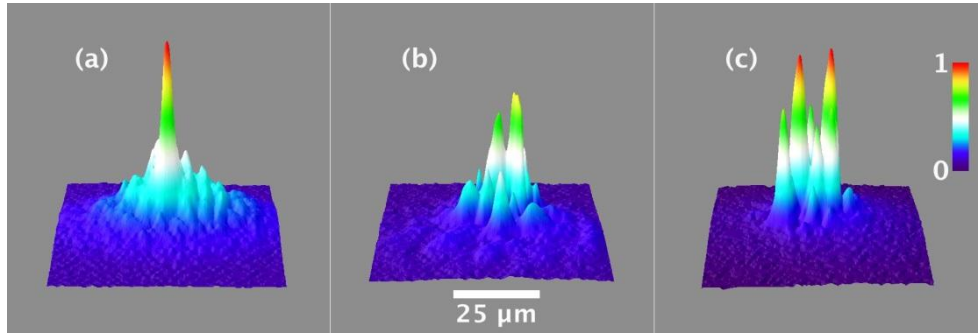

**Figure 6:** Experimental results (intensity profiles of the probe beam around a RW event) on the nonlinear propagation of an intense femtosecond probe beam in the bulk of a glass with a LHs lattice inscribed in it. (a) Under the effect of Kerr self-focusing at the limit of the critical power a linear RW is further amplified maintaining its contrast from the surrounding waves. (b) As the input power and nonlinearities are increased one can observe the appearance of small scale multifilaments. (c) At even higher input powers the multifilaments shown in (b) reach the clamping intensity (red peaks) and thus the RW statistics are destroyed.

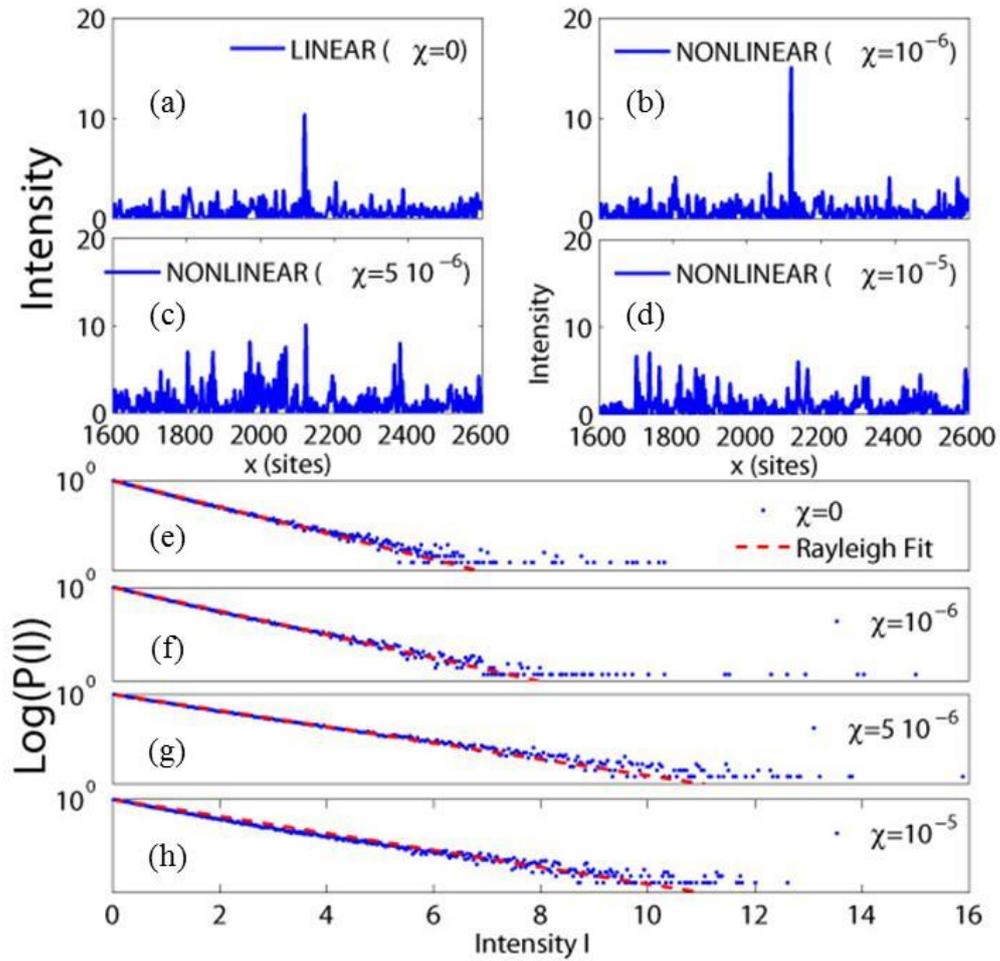

**Figure 7:** Simulations on the role of the nonlinearity. (a) An intensity profile in the linear regime of a region presenting a RW surrounded by low amplitude waves, (b) the same region in the nonlinear regime (at the limit of the critical power) showing an increase of the amplitude of the related waves maintaining though a clear RW picture. As the nonlinearity increases though (c-d) more waves are amplified destroying the rogue nature. (e-h) Intensity distributions of the (a-d) cases respectively, again showing that at high nonlinearities the rogue statistics are destroyed.

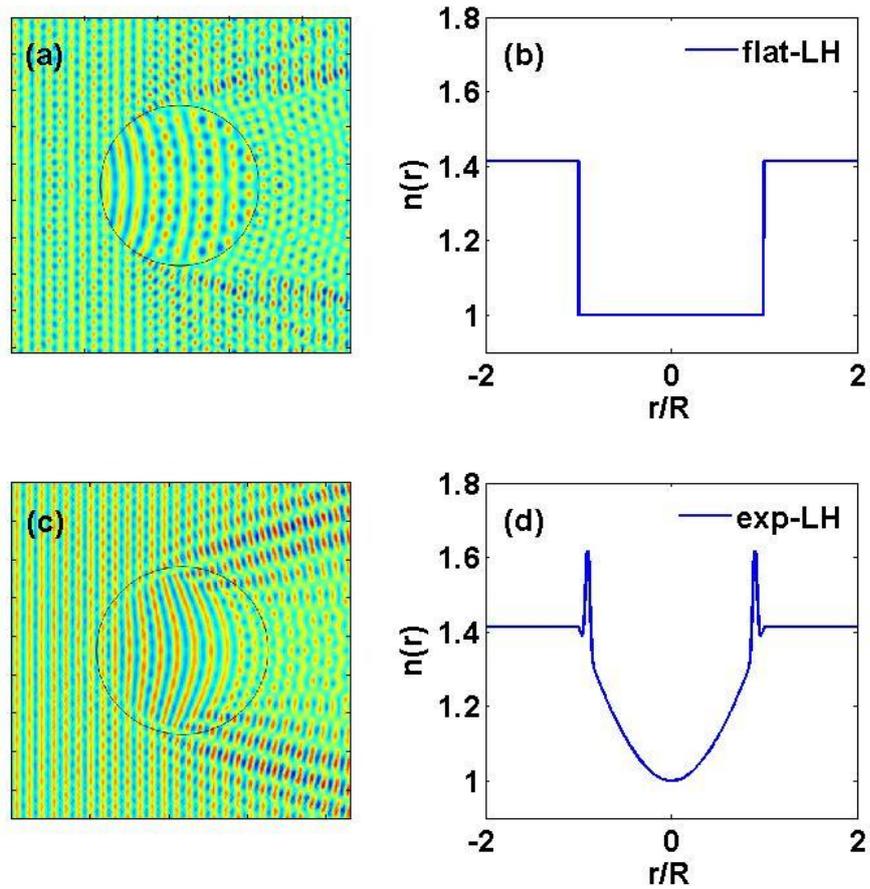

**Figure 8:** (a,c) EM plane wave propagation simulations and (b,d) the characteristic refractive index profile of a flat index profile LH (flat-LH) and of a LH presenting an index profile close to the experimental one (exp-LH) respectively.

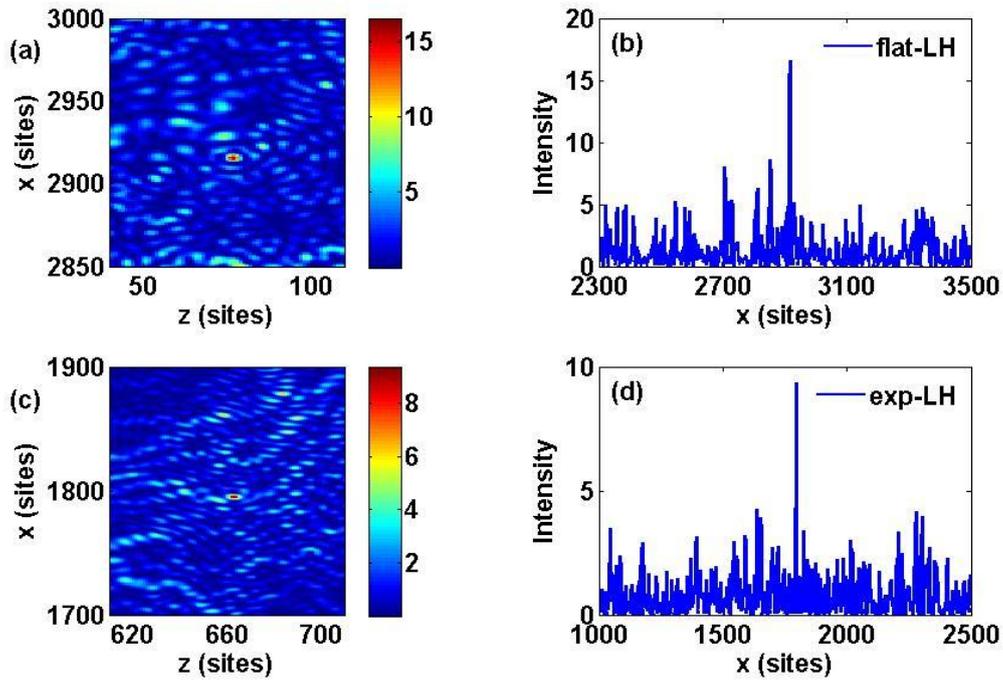

**Figure 9:** Simulation results of a monochromatic plane wave propagation in a 2D random network and corresponding intensity profiles of RW regions using (a,b) flat-LHs and (c,d) exp-LHs. In both cases we observe RW events as was the case with regular LH lattices generalizing thus our findings independently of the exact type of scatterers.

18